\documentclass[proceedings]{JHEP37}
\usepackage{amsfonts}
\usepackage{amsmath}
\usepackage{epsfig}

\newbox\mybox

\newcommand\fverb{\setbox\mybox=\hbox\bgroup\verb}
\newcommand\fverbdo{\egroup\medskip\noindent\fbox{\unhbox\mybox}\ }
\newcommand\fverbit{\egroup\item[\fbox{\unhbox\mybox}]}
\conference{Non-Hermitian systems of Euclidean type}
\abstract{We study several classes of non-Hermitian Hamiltonian systems,
which can be expressed in terms of bilinear combinations of 
Euclidean Lie algebraic generators. The classes are distinguished by
different versions of antilinear (PT)-symmetries exhibiting various types
of qualitative behaviour. On the basis of explicitly computed non-perturbative
Dyson maps we construct metric operators, isospectral Hermitian counterparts for which we
solve the corresponding time-independent Schr\"{o}dinger equation for
specific choices of the coupling constants. In these cases general analytical expressions 
for the solutions are obtained in the form of Mathieu functions, which we analyze numerically
to obtain the corresponding energy eigenspectra. We identify regions in the
parameter space for which the corresponding spectra are entirely real and
also domains where the PT symmetry is spontaneously broken and sometimes also regained 
at exceptional points. In some cases it is shown explicitly how the threshold region from real to complex 
spectra is characterized by the breakdown of the Dyson maps or the metric operator. We establish the
explicit relationship to models currently under investigation in the context of beam dynamics in optical
lattices.}

\title{Non-Hermitian systems of Euclidean Lie algebraic type with real
eigenvalue spectra}
\author{Sanjib Dey, Andreas Fring and Thilagarajah Mathanaranjan \\
Department of Mathematical Science, City University London,\\
Northampton Square, London EC1V 0HB, UK\\
E-mail: sanjib.dey.1@city.ac.uk, a.fring@city.ac.uk,
thilagarajah.mathanaranjan.1@city.ac.uk}

\input{tcilatex}
\begin{document}

\section{Introduction}

Quasi-exactly solvable models \cite{Tur0} of Lie algebraic type are believed
to be almost all related to $sl_{2}(\mathbb{C})$ with their compact and
non-compact real forms $su(2)$ and $su(1,1)$, respectively \cite{Hum}.\ The
nature of those models dictates that essentially all the wavefunctions
related to solutions for the time-independent Schr\"{o}dinger equation of
these type of models may be expressed in terms of hypergeometric functions.
Non-Hermitian variants of these models expressed generically in terms of $%
su(2)$ or $su(1,1)$ generators have been investigated systematically in \cite%
{PEGAAF2,Paulos} and large classes of models were found to possess real or
partially spectra despite their non-Hermitian nature. Under certain
constraints on the coupling constants the models could be mapped to
Hermitian isospectral counterparts. Positive Hermitian metric operators were
shown to exist, such that a consistent quantum mechanical description of
these models is possible when following the general techniques developed
over the last years \cite{Bender:1998ke,Benderrev,Alirev} in the context of $%
\mathcal{PT}$-symmetric non-Hermitian quantum mechanics.

It is, however, also well known that there exists an interesting subclass of
solvable models related to Mathieu functions which are known to possess
solutions, which are not expressible in terms of hypergeometric functions.
In a more generic setting these type of models are known to be related to
specific representations of the Euclidean algebra rather than to its
subalgebra $sl_{2}(\mathbb{C})$. This feature makes models based on them
interesting objects of investigation from a mathematical point of view. In a
more applied setting it is also well known that the Mathieu equation arises
in optics as a reduction from the Helmholtz equation. This analogue setting
of complex quantum mechanics is currently under intense investigation.
Concrete versions of complex potentials leading to real Mathieu potentials
have recently been studied from a theoretical as well as experimental point
of view in \cite{Muss,MatMakris,Guo,OPMidya,MatHugh,MatHughEva}. Further
applications are found for instance in the investigation of complex crystals 
\cite{MatLongo}.

It was recently shown that for $E_{2}$ \cite{BenKal} and $E_{3}$ \cite%
{JonSmKal} some simple non-Hermitian versions also possess real spectra.
Here we will follow the line of thought of \cite{PEGAAF2} and investigate
systematically the analogues of quasi-exactly solvable models of Lie
algebraic type, that is those models which can be written as bilinear
combinations in terms of the Euclidean algebra generators.

Our manuscript is organized as follows: At the beginning of section 2 we
discuss five different types of $\mathcal{PT}$-symmetries for the $E_{2}$%
-algebra and present the computation of the adjoint action on their
generators. In the following five subsection we derive Dyson maps and
isospectral counterparts for generic non-Hermitian Hamiltonians invariant
under these different types of symmetries. For the last symmetry we present
a more detailed analysis of the time-independent Schr\"{o}dinger equation.
We derive some explicit analytical solutions, which we analyze numerically\
to compute the corresponding energy eigenspectra leading to three
qualitatively different scenarios: entirely real eigenvalue spectra, spectra
with spontaneously broken $\mathcal{PT}$-symmetry at exceptional points
characterized by two or three disconnected regions in the parameter space.
In section 3 we discuss the $\mathcal{PT}$-symmetries for the $E_{3}$%
-algebra, present the computation of the adjoint action on its generators
and indicate how to obtain simple examples of explicit isospectral pairs of
an $E_{3}$-invariant non-Hermitian and Hermitian Hamiltonian.

\section{$\mathcal{PT}$-symmetric E$_{2}$-invariant non-Hermitian
Hamiltonians}

We take here the commutation relations obeyed by the three generators $u$,$v$
and $J$ as the defining relations of the Euclidean-algebra $E_{2}$ 
\begin{equation}
\left[ u,J\right] =iv,\qquad \left[ v,J\right] =-iu,\qquad \text{and\qquad }%
\left[ u,v\right] =0.  \label{E2}
\end{equation}%
Obviously there are many representations for this algebra, as for instance
one used in the context of quantizing strings on tori \cite{Isham} acting on
square integrable wavefunctions $L^{2}(\mathbf{S}^{1},d\theta )$ with%
\begin{equation}
J:=-i\partial _{\theta },\qquad u:=\sin \theta ,\qquad \text{and\qquad }%
v:=\cos \theta ,  \label{rep1}
\end{equation}%
or a two-dimensional one in terms of generators of the Heisenberg canonical
commutators $q_{j}$, $p_{j}$ satisfying $\left[ q_{j},p_{k}\right] =i\delta
_{jk}$ for $j,k=1,2$%
\begin{equation}
J:=q_{1}p_{2}-p_{1}q_{2},\qquad u:=p_{2},\qquad \text{and\qquad }v:=p_{1}.
\label{rep2}
\end{equation}%
For our purposes it is important to note that the $E_{2}$-algebra is left
invariant with regard to an antilinear symmetry \cite{EW}. As previously
noted \cite{Bender:2002yp,Mon1,DFG} in dimensions larger than one there are
in general various types of antilinear symmetries, which by a slight abuse
of language we all refer to as $\mathcal{PT}$-symmetries. For instance, it
is easy to see that the algebra (\ref{E2}) is left invariant under the
following antilinear maps%
\begin{equation}
\begin{array}{lllll}
\mathcal{PT}_{1}:~~~~~ & J\rightarrow -J,~~ & u\rightarrow -u,~~ & 
v\rightarrow -v,~~~ & i\rightarrow -i, \\ 
\mathcal{PT}_{2}: & J\rightarrow -J, & u\rightarrow u, & v\rightarrow v, & 
i\rightarrow -i, \\ 
\mathcal{PT}_{3}: & J\rightarrow J, & u\rightarrow v, & v\rightarrow u, & 
i\rightarrow -i, \\ 
\mathcal{PT}_{4}: & J\rightarrow J, & u\rightarrow -u, & v\rightarrow v, & 
i\rightarrow -i, \\ 
\mathcal{PT}_{5}: & J\rightarrow J, & u\rightarrow u, & v\rightarrow -v, & 
i\rightarrow -i.%
\end{array}
\label{PT}
\end{equation}%
Each of these symmetries may be utilized to describe different types of
physical scenarios. For instance, $\mathcal{PT}_{1}$ was considered in \cite%
{BenKal} with $\mathcal{P}_{1}:\theta \rightarrow \theta +\pi $
corresponding to a reflection of the particle to the opposite side of the
circle for the representation (\ref{rep1}). For the same representation we
can identify the remaining symmetries as $\mathcal{P}_{2}:\theta \rightarrow
\theta +2\pi $, $\mathcal{P}_{3}:\theta \rightarrow \pi /2-\theta $, $%
\mathcal{P}_{4}:\theta \rightarrow \pi -\theta $ and $\mathcal{P}_{5}:\theta
\rightarrow -\theta $. Of course other representations allow for different
interpretations. For instance, in the two dimensional representation (\ref%
{rep2}) the symmetry $\mathcal{PT}_{3}$ can be used when describing systems
with two particle species as it may be viewed as a particle exchange, or an
annihilation of a particle of one species accompanied by the creation a
particle of another species, together with a simultaneous reflection $%
\mathcal{PT}_{3}:p_{1}\leftrightarrow p_{2},$ $q_{1}\leftrightarrow -q_{2}$.

$\mathcal{PT}_{i}$-invariant Hamiltonians $H$ in term of bilinear
combinations of $E_{2}$-generators are then easily written down. Crucially,
this very general symmetry allows for non-Hermitian Hamiltonians to be
considered since it is antilinear \cite{EW}. Following the general
techniques developed over the last years \cite%
{Bender:1998ke,Benderrev,Alirev} in the context of $\mathcal{PT}$-symmetric
non-Hermitian quantum mechanics we attempt to map these non-Hermitian
Hamiltonians $H\neq H^{\dagger }$ to isospectral Hermitian counterparts $%
h=h^{\dagger }$ by means of a similarity transformation $h=\eta H\eta ^{-1}$%
. When $\eta $, often referred to as the Dyson map, is Hermitian the latter
equation is equivalent to $H^{\dagger }=\eta ^{2}H\eta ^{-2}$, which is
another equation one might utilize to determine $\eta $. Taking here the
Dyson map to be of the general form 
\begin{equation}
\eta =e^{\lambda J+\rho u+\tau v},\qquad \text{\ \ \ \ \ \ \ \ for }\lambda
,\tau ,\rho \in \mathbb{R},  \label{eta}
\end{equation}%
we can easily compute the adjoint action of this operator on the $E_{2}$%
-generators. We find%
\begin{eqnarray}
\eta J\eta ^{-1} &=&J+i(\rho v-\tau u)\frac{\sinh \lambda }{\lambda }+(\rho
u+\tau v)\frac{1-\cosh \lambda }{\lambda },  \label{ad1} \\
\eta u\eta ^{-1} &=&u\cosh \lambda -iv\sinh \lambda ,  \label{ad2} \\
\eta v\eta ^{-1} &=&v\cosh \lambda +iu\sinh \lambda .  \label{ad3}
\end{eqnarray}%
Once $\eta $ is identified the metric operators needed for a consistent
quantum mechanical formulation can in general be taken to be $\rho =\eta
^{\dagger }\eta $. Let us now construct isospectral counterparts, if they
exist, for non-Hermitian Hamiltonians symmetric with regard to the various
different types of $\mathcal{PT}$-symmetries. It should be noted that exact
computations of this type remain a rare exception and even for some of the
simplest potentials the answer is only known perturbatively, as for instance
even for the simple prototype non-Hermitian potential $V=i\varepsilon x^{3}$ 
\cite{Bender:2004sa,Mostafazadeh:2004qh,CA}.

\subsection{$\mathcal{PT}_{1}$-invariant Hamiltonians of $E_{2}$-Lie
algebraic type}

The most general $\mathcal{PT}_{1}$-invariant Hamiltonian expressed in terms
of bilinear combinations of the $E_{2}$-generators is 
\begin{equation}
H_{\mathcal{PT}_{1}}=\mu _{1}J^{2}+i\mu _{2}J+i\mu _{3}u+i\mu _{4}v+\mu
_{5}u\!J+\mu _{6}vJ+\mu _{7}u^{2}+\mu _{8}v^{2}+\mu _{9}uv,  \label{HPT1}
\end{equation}%
with $\mu _{i}\in \mathbb{R}$ for $i=1,\ldots ,9$. Clearly the Hamiltonian $%
H_{\mathcal{PT}_{1}}$ is non-Hermitian with regard to the standard inner
product when considering it for a Hermitian representation with $J^{\dagger
}=J$, $v^{\dagger }=v$ and $u^{\dagger }=u$, unless $\mu _{2}=0$, $\mu
_{5}=-2\mu _{4}$, $\mu _{6}=2\mu _{3}$. The specific case $H_{BK}=J^{2}+igv$
when $\mu _{i}=0$ for $i\neq 1,4$ was studied in \cite{BenKal}, where
partially real spectra were found but no isospectral counterparts were
constructed. Using the relations (\ref{ad1})-(\ref{ad3}), we compute the
adjoint action of $\eta $ on $H$ and subsequently demand the result to be
Hermitian. This requirement will constrain our 12 free parameters $\mu
_{i},\lambda ,\tau ,\rho $. A priori it is unclear whether solutions to the
resulting set of equations exist. For $H_{\mathcal{PT}_{1}}$ we find the
manifestly Hermitian isospectral counterpart%
\begin{equation}
h_{\mathcal{PT}_{1}}=\mu _{1}J^{2}+\mu _{3}\{v,J\}-\mu _{4}\{u,J\}-\frac{%
2\mu _{3}\mu _{4}}{\mu _{1}}uv+\frac{\mu _{4}^{2}-\mu _{3}^{2}}{\mu _{1}}%
u^{2}+\mu _{8}(u^{2}+v^{2}).
\end{equation}%
As usual, we denote by $\{A,B\}:=AB+BA$ the anti-commutator. Without loss of
generality we may set $\mu _{8}=0$ since $C=u^{2}+v^{2}$ is a Casimir
operator for the $E_{2}$-algebra and can therefore always be added to $H$ \
having simply the effect of shifting the ground state energy. The remaining
constants $\mu _{i}$ have been constrained to 
\begin{equation}
\tau =\frac{\lambda \mu _{3}}{\mu _{1}},~\rho =-\frac{\lambda \mu _{4}}{\mu
_{1}},~\mu _{2}=0,~\mu _{5}=-2\mu _{4},~\mu _{6}=2\mu _{3},~\mu _{7}=\mu
_{8}+\frac{\mu _{4}^{2}-\mu _{3}^{2}}{\mu _{1}},~\mu _{9}=-\frac{2\mu
_{3}\mu _{4}}{\mu _{1}},~  \label{con}
\end{equation}%
by the requirement that $h_{\mathcal{PT}_{1}}$ ought to be Hermitian,
whereas $\lambda ,\mu _{1},\mu _{3},\mu _{4}$ are chosen to be free. We
observe that we have been led to the constraints (\ref{con}), of which a
subset stated that $H_{\mathcal{PT}_{1}}$ is already Hermitian before the
transformation. We also note that the constraints (\ref{con}) do not allow a
reduction to the Hamiltonian $H_{BK}$, dealt with in \cite{BenKal}, as for
instance $\mu _{5}=0$ implies $\mu _{4}=0$.

Having guaranteed that $H_{\mathcal{PT}_{1}}$ possess real eigenvalues under
certain constraints we may now also compute the corresponding solutions to
the time-independent Schr\"{o}dinger equation $h_{\mathcal{PT}_{1}}\phi
=E\phi $ or equivalently to $H_{\mathcal{PT}_{1}}\psi =E\psi $ with $\psi
=\eta ^{-1}\phi $. We find 
\begin{equation}
\phi (\theta )=e^{-\frac{i\mu _{4}\cos \theta }{\mu _{1}}-i\frac{\sin \theta 
}{\mu _{1}}\mu _{3}}\left[ c_{1}\exp \left( -i\theta \sqrt{\frac{E}{\mu _{1}}%
+\frac{\mu _{3}^{2}}{\mu _{1}^{2}}}\right) +\frac{i}{2\sqrt{\frac{E}{\mu _{1}%
}+\frac{\mu _{3}^{2}}{\mu _{1}^{2}}}}c_{2}\exp \left( i\theta \sqrt{\frac{E}{%
\mu _{1}}+\frac{\mu _{3}^{2}}{\mu _{1}^{2}}}\right) \right] ,
\end{equation}%
with normalization constants $c_{1}$, $c_{2}$. Imposing either bosonic or
fermionic boundary conditions, i.e. $\psi (\theta +2\pi )=\pm \psi (\theta )$%
, we obtain the discrete real energy eigenvalues%
\begin{equation}
\text{bosonic:~~}E_{n}=\mu _{1}\left( n^{2}-\frac{\mu _{3}^{2}}{\mu _{1}^{2}}%
\right) ,\qquad \text{fermionic:~~}E_{n}=\mu _{1}\left( n^{2}+n+\frac{1}{4}-%
\frac{\mu _{3}^{2}}{\mu _{1}^{2}}\right) ,~~n\in \mathbb{Z}\text{.}
\end{equation}%
As expected, the wavefunctions are eigenstates of the $\mathcal{PT}$%
-operator, selecting different behaviours for the two linearly independent
parts of $\phi (\theta )$, acting as $\mathcal{PT}_{1}\phi
_{n}(c_{1})=(-1)^{n}\phi _{n}(c_{1})$ and $\mathcal{PT}_{1}\phi
_{n}(c_{2})=(-1)^{n+1}\phi _{n}(c_{2})$.

\subsection{$\mathcal{PT}_{2}$-invariant Hamiltonians of $E_{2}$-Lie
algebraic type}

Similarly as in the previous subsection we use the adjoint action of $\eta $
as specified in (\ref{eta}) to map the general $\mathcal{PT}_{2}$-symmetric
and for $\mu _{2}\neq 0$, $\mu _{5}\neq 2\mu _{4}$, $\mu _{6}=-2\mu _{3}$
non-Hermitian Hamiltonian%
\begin{equation}
H_{\mathcal{PT}_{2}}=\mu _{1}J^{2}+i\mu _{2}J+\mu _{3}u+\mu _{4}v+i\mu
_{5}u~\!\!J+i\mu _{6}v~\!\!J+\mu _{7}u^{2}+\mu _{8}v^{2}+\mu _{9}uv,
\end{equation}%
to the Hermitian isospectral counterpart%
\begin{eqnarray}
h_{\mathcal{PT}_{2}} &=&\mu _{1}J^{2}+\mu _{3}\tanh \frac{\lambda }{2}%
\{u,J\}+\mu _{4}\tanh \frac{\lambda }{2}\{u,J\}+\frac{2\mu _{3}\mu _{4}}{\mu
_{1}}\tanh ^{2}\frac{\lambda }{2}uv \\
&&+\frac{\mu _{3}^{2}}{\mu _{1}}\frac{\cosh \lambda }{\cosh ^{2}\frac{%
\lambda }{2}}u^{2}+\left( \frac{\mu _{3}^{2}}{\mu _{1}}+\frac{\mu _{4}^{2}}{%
\mu _{1}}\tanh ^{2}\frac{\lambda }{2}\right) v^{2}+\mu _{8}(u^{2}+v^{2}). 
\notag
\end{eqnarray}%
In this case the coupling constants are constraint to

\begin{equation}
\rho =\tau \frac{\mu _{3}}{\mu _{4}}=\frac{\mu _{3}\lambda \coth \lambda }{%
\mu _{1}},~~\mu _{2}=0,~~\mu _{5}=2\mu _{4},~\mu _{6}=-2\mu _{3},~~\mu
_{7}=\mu _{8}+\frac{\mu _{3}^{2}-\mu _{4}^{2}}{\mu _{1}},~\mu _{9}=\frac{%
2\mu _{3}\mu _{4}}{\mu _{1}},
\end{equation}%
We note that once again we have only the four free parameters $\lambda ,\mu
_{1},\mu _{3},\mu _{4}$ left at our disposal, as $\mu _{8}$ may be set to
zero for the above mentioned reason. As in the previous case these
conditions imply also that the original Hamiltonian $H_{\mathcal{PT}_{2}}$
is already Hermitian when these type of constraints are imposed.

\subsection{$\mathcal{PT}_{3}$-invariant Hamiltonians of $E_{2}$-Lie
algebraic type}

As the general $\mathcal{PT}_{3}$-invariant Hamiltonian of Lie algebraic
type we consider%
\begin{eqnarray}
H_{\mathcal{PT}_{3}} &=&\mu _{1}J^{2}+\mu _{2}J+\mu _{3}(u+v)+i\mu
_{4}(u-v)+\mu _{5}(u+v)~\!\!J+i\mu _{6}(u-v)~\!\!J+i\mu _{7}(v^{2}-u^{2}) 
\notag \\
&&+\mu _{8}(v^{2}+u^{2})+\mu _{9}uv.
\end{eqnarray}%
For Hermitian representations of the $E_{2}$-generators this Hamiltonian is
non-Hermitian unless $\mu _{6}=\mu _{7}=0$ and $\mu _{5}=2\mu _{4}$. As
isospectral Hermitian counterpart we find in this case 
\begin{eqnarray}
h_{\mathcal{PT}_{3}} &=&\mu _{1}J^{2}+\mu _{2}J+\frac{1}{2}\left( \mu
_{5}+\mu _{6}\tanh \frac{\lambda }{2}\right) \{u+v,J\} \\
&&\!\!\!\!\!\!\!\!\!\!\!\!+\left\{ \frac{1}{2\mu _{1}}\left[ \mu
_{5}^{2}+\mu _{6}^{2}\tanh ^{2}\frac{\lambda }{2}+\mu _{6}\mu _{5}\frac{%
4+4\cosh \lambda -2\cosh (2\lambda )}{\sinh (2\lambda )}\right] +\frac{2\mu
_{7}}{\sinh (2\lambda )}\right\} uv  \notag \\
&&\!\!\!\!\!\!\!\!\!\!\!\!+\left[ \mu _{3}-\frac{\mu _{6}}{2}+\left( \mu
_{4}-\frac{\mu _{5}}{2}\right) \tanh \frac{\lambda }{2}\right] (u+v)+\left[
\mu _{8}+\frac{\mu _{5}\mu _{6}\sinh \lambda +\mu _{6}^{2}\cosh \lambda }{%
2\mu _{1}(1+\cosh \lambda )}\right] (u^{2}+v^{2})  \notag
\end{eqnarray}%
with only four constraining equations%
\begin{eqnarray}
\rho &=&\tau =\frac{\lambda \left( \mu _{5}+\mu _{6}\coth \lambda \right) }{%
2\mu _{1}},~~~\coth \lambda =\frac{\mu _{2}\mu _{5}+\mu _{1}\left( \mu
_{6}-2\mu _{3}\right) }{\mu _{1}\left( 2\mu _{4}-\mu _{5}\right) -\mu
_{2}\mu _{6}},~  \label{co1} \\
\mu _{9} &=&\frac{\mu _{5}^{2}+\mu _{6}^{2}+2\mu _{6}\mu _{5}\coth (2\lambda
)}{2\mu _{1}}+2\mu _{7}\coth (2\lambda ).
\end{eqnarray}%
Thus, in this case we have eight free parameters left. We also note that
unlike as for the $\mathcal{PT}_{1}$ and $\mathcal{PT}_{2}$ symmetric cases
we are not led to constraints which render the original Hamiltonian $H_{%
\mathcal{PT}_{3}}$ Hermitian. For $\mu _{1}=1$, $\mu _{7}=2q$ and all other
coupling constants vanishing the Schr\"{o}dinger equation with
representation (\ref{rep1}) converts into the standard Mathieu differential
equation, see e.g. \cite{Grad},%
\begin{equation}
-\phi ^{\prime \prime }(\theta )+2iq\cos (2\theta )\phi (\theta )=E\phi
(\theta ).  \label{mat}
\end{equation}%
with purely complex coupling constant. Unfortunately for this choice of the
coupling constants the Dyson map is no longer well-defined, because of the
last equation in (\ref{co1}), such that it remains an open problem to find
the corresponding isospectral counterpart for this scenario.

\subsection{$\mathcal{PT}_{4}$-invariant Hamiltonians of $E_{2}$-Lie
algebraic type}

The general $\mathcal{PT}_{4}$-invariant Hamiltonian we consider is%
\begin{equation}
H_{\mathcal{PT}_{4}}=\mu _{1}J^{2}+\mu _{2}J+i\mu _{3}u+\mu _{4}v+i\mu
_{5}u~\!\!J+\mu _{6}v~\!\!J+\mu _{7}u^{2}+\mu _{8}v^{2}+i\mu _{9}uv.
\end{equation}%
This Hamiltonian is non-Hermitian unless $\mu _{5}=\mu _{9}=0$ and $\mu
_{6}=2\mu _{3}$. Constraining now the parameters as 
\begin{eqnarray}
&&\rho =0,\quad \tau =\frac{\lambda \left( \mu _{5}\coth \lambda +\mu
_{6}\right) }{2\mu _{1}},\quad \coth (2\lambda )=\frac{4\mu _{1}(\mu
_{8}-\mu _{7})-\mu _{5}^{2}-\mu _{6}^{2}}{2\mu _{5}\mu _{6}}, \\
&&\mu _{3}=\frac{\mu _{1}\mu _{5}+\mu _{2}\mu _{6}-2\mu _{1}\mu _{4}}{2\mu
_{1}}\tanh \lambda +\frac{\mu _{2}\mu _{5}}{2\mu _{1}}+\frac{\mu _{6}}{2}%
,\qquad \mu _{9}=0,\quad
\end{eqnarray}%
we map this to the isospectral counterpart 
\begin{eqnarray}
h_{\mathcal{PT}_{4}} &=&\mu _{1}J^{2}+\mu _{2}J+\frac{1}{2}\left( \mu
_{6}+\mu _{5}\tanh \frac{\lambda }{2}\right) \{v,J\} \\
&&+\left[ \frac{\mu _{2}\tanh \left( \frac{\lambda }{2}\right) \left( \mu
_{5}+\mu _{6}\tanh \lambda \right) }{2\mu _{1}}+\left( \mu _{4}-\frac{\mu
_{5}}{2}\right) \func{sech}\lambda \right] v  \notag \\
&&+\left[ \frac{\mu _{5}^{2}\left( \tanh ^{2}\frac{\lambda }{2}-\cosh
(2\lambda )\right) -2\mu _{6}^{2}\sinh ^{2}\lambda +2\mu _{5}\mu _{6}\left(
\tanh \frac{\lambda }{2}-\sinh (2\lambda )\right) }{8\mu _{1}}\right.  \notag
\\
&&+\left. \frac{\mu _{8}-\mu _{7}}{2}\cosh (2\lambda )\right] \left(
v^{2}-u^{2}\right) +\frac{\mu _{5}^{2}\cosh \lambda +\mu _{5}\mu _{6}\sinh
\lambda }{4\mu _{1}(1+\cosh \lambda )}+\frac{1}{2}\left( \mu _{7}+\mu
_{8}\right) .  \notag
\end{eqnarray}%
Thus, in this case we have seven free parameters left to our disposal. Also
in this case we obtained a genuine non-Hermitian/Hermitian isospectral pair
of Hamiltonians.

\subsection{$\mathcal{PT}_{5}$-invariant Hamiltonians of $E_{2}$-Lie
algebraic type}

As general $\mathcal{PT}_{5}$-invariant Hamiltonian we consider 
\begin{equation}
H_{\mathcal{PT}5}=\mu _{1}J^{2}+\mu _{2}J+\mu _{3}u+i\mu _{4}v+\mu
_{5}u~\!\!J+i\mu _{6}v~\!\!J+\mu _{7}u^{2}+\mu _{8}v^{2}+i\mu _{9}uv.~~
\end{equation}%
This Hamiltonian is non-Hermitian unless $\mu _{6}=\mu _{9}=0$ and $\mu
_{5}=-2\mu _{4}$. In the same manner as in the previous subsections we
construct the isospectral counterpart 
\begin{eqnarray}
h_{\mathcal{PT}_{5}} &=&\mu _{1}J^{2}+\mu _{2}J+\frac{1}{2}\left( \mu
_{5}-\mu _{6}\tanh \frac{\lambda }{2}\right) \{u,J\} \\
&&+\left[ \frac{2\mu _{5}^{2}\sinh ^{2}\lambda +\mu _{6}^{2}(\func{sech}^{2}%
\frac{\lambda }{2}+\cosh (2\lambda )-1)+2(\tanh \frac{\lambda }{2}-\sinh
(2\lambda ))\mu _{5}\mu _{6}}{8\mu _{1}}\right.  \notag \\
&&+\left. \frac{\mu _{8}-\mu _{7}}{2}\cosh (2\lambda )\right] (v^{2}-u^{2})+%
\left[ \func{csch}\lambda \left( \mu _{4}+\frac{1}{2}\mu _{5}\right) +\frac{%
\mu _{2}}{2\mu _{1}}(\mu _{5}-\coth \lambda \mu _{6})\right] u  \notag \\
&&+\frac{\mu _{6}^{2}\cosh \lambda -\mu _{5}\mu _{6}\sinh \lambda }{4\mu
_{1}(1+\cosh \lambda )}+\frac{1}{2}\left( \mu _{7}+\mu _{8}\right) ,  \notag
\end{eqnarray}%
where the constants are constraint to$\allowbreak $%
\begin{eqnarray}
&&\tau =0,\quad \rho =\frac{\lambda \left( \mu _{5}-\mu _{6}\coth \lambda
\right) }{2\mu _{1}},\quad \coth (2\lambda )=\frac{\mu _{5}^{2}+\mu
_{6}^{2}-4\mu _{1}\mu _{7}+4\mu _{1}\mu _{8}}{2\mu _{5}\mu _{6}}, \\
&&\mu _{3}=\frac{\left( 2\mu _{1}\mu _{4}+\mu _{1}\mu _{5}-\mu _{2}\mu
_{6}\right) \coth (\lambda )}{2\mu _{1}}+\frac{\mu _{2}\mu _{5}}{2\mu _{1}}-%
\frac{\mu _{6}}{2},\quad \mu _{9}=0.\quad
\end{eqnarray}%
Thus, in this case we have also seven free parameters left to our disposal.

Having obtained the Hermitian counterpart, let us construct in this case
some solutions to the time-independent Schr\"{o}dinger equation. The
discussion of the entire parameter space is a formidable task, but as we
shall see it will be sufficient to focus on some special parameter choices
in order to extract different types of qualitative behaviour. We will also
make contact to some special cases previously treated in the literature,
notably in the area of complex optical lattices.

\subsubsection{Maps to a three parameter real Mathieu equation}

First we specify our parameters further such that only three are left free 
\begin{eqnarray}
\mu _{1} &=&1,\quad \mu _{2}=0,\quad \mu _{5}=-2\mu _{4},\quad \mu
_{6}=-2\mu _{3},\quad \mu _{8}=\mu _{9}=0,\quad  \label{cons1} \\
\tau &=&0,\quad \rho =\lambda \left( \mu _{3}\coth \lambda -\mu _{4}\right)
,\quad \coth (2\lambda )=\frac{\mu _{3}^{2}+\mu _{4}^{2}-\mu _{7}}{2\mu
_{3}\mu _{4}}.  \label{cons2}
\end{eqnarray}%
The corresponding isospectral pair of Hamiltonians simplifies in this case
to 
\begin{eqnarray}
H_{\mathcal{PT}_{5}}^{(3)} &=&J^{2}-i\mu _{3}\{v,~\!\!J\}-\mu
_{4}\{u,~\!\!J\}+\mu _{7}u^{2},~~ \\
h_{\mathcal{PT}_{5}}^{(3)} &=&J^{2}+\alpha \{u,J\}+\beta u^{2}+\gamma ,
\end{eqnarray}%
where $\alpha $, $\beta $, $\gamma $ are functions of $\mu _{3}$, $\mu _{4}$%
, $\mu _{7}$%
\begin{eqnarray}
\alpha &=&\mu _{3}\tanh \frac{\lambda }{2}-\mu _{4}, \\
\beta &=&\frac{2\mu _{3}}{1+\cosh \lambda }(\mu _{3}\cosh \lambda -\mu
_{4}\sinh \lambda )+\mu _{7}-2\gamma , \\
\gamma &=&(\mu _{3}\cosh \lambda -\mu _{4}\sinh \lambda )^{2}-\mu _{7}\sinh
^{2}\lambda .
\end{eqnarray}%
For the representation (\ref{rep1}) the standard Mathieu differential
equation (\ref{mat}) with real coupling constant is easily converted into
the time-independent Schr\"{o}dinger equation 
\begin{equation}
h_{\mathcal{PT}_{5}}^{(3)}\psi (\theta )=E\psi (\theta )  \label{S4}
\end{equation}%
with the transformations $\phi (\theta )\rightarrow e^{-i\alpha \cos \theta
}\psi (\theta )$, $q\rightarrow (\alpha ^{2}-\beta )/4$ and $E\rightarrow
E+(\alpha ^{2}-\beta )/2-\gamma $. Therefore (\ref{S4}) is solved by%
\begin{equation}
\psi (\theta )=e^{i\alpha \cos \theta }\left[ c_{1}C\left( E+\frac{\alpha
^{2}-\beta }{2}-\gamma ,\frac{\alpha ^{2}-\beta }{4},\theta \right)
+c_{2}S\left( E+\frac{\alpha ^{2}-\beta }{2}-\gamma ,\frac{\alpha ^{2}-\beta 
}{4},\theta \right) \right]
\end{equation}%
where $C$ and $S$ denote the even and odd Mathieu function, respectively. A
discrete energy spectrum is extracted in the usual way by imposing periodic
boundaries $\psi (\theta +2\pi )=e^{i\pi s}\psi (\theta )$ as quantization
condition. While in general anyonic conditions are possible in dimensions
lower than 4, we present here only the bosonic and fermionic case, that is $%
s=0$ and $s=1$, respectively. As the Mathieu function is known to possess
infinitely many periodic solutions, the boundary condition as such is not
sufficient to obtain a unique solution. However, the latter is achieved by
demanding in addition the continuity of the energy levels at $q=0$. The
inclusion of all values for $s$ will naturally lead to band structures.

\begin{figure}[h]
\centering   \includegraphics[width=7.5cm,height=6.0cm]{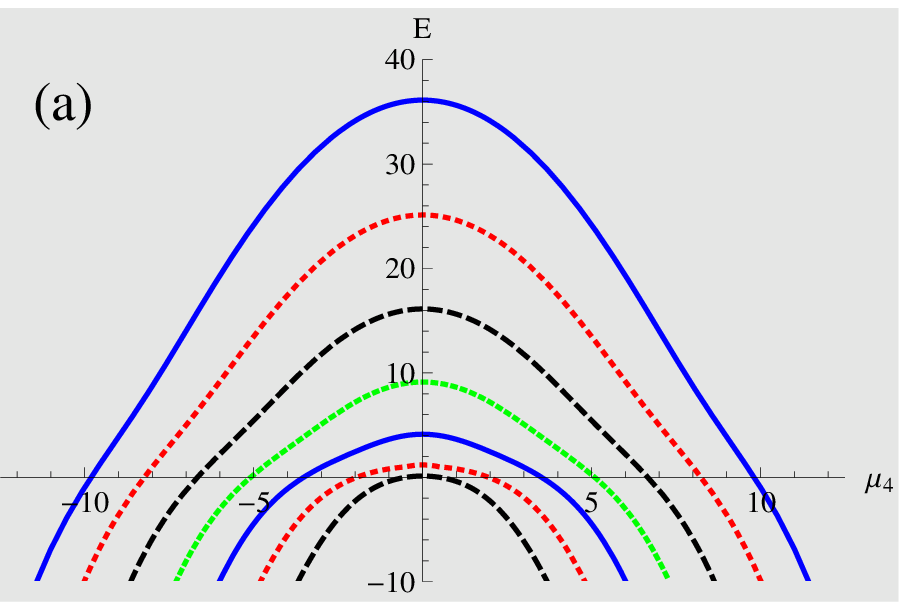} %
\includegraphics[width=7.5cm,height=6.0cm]{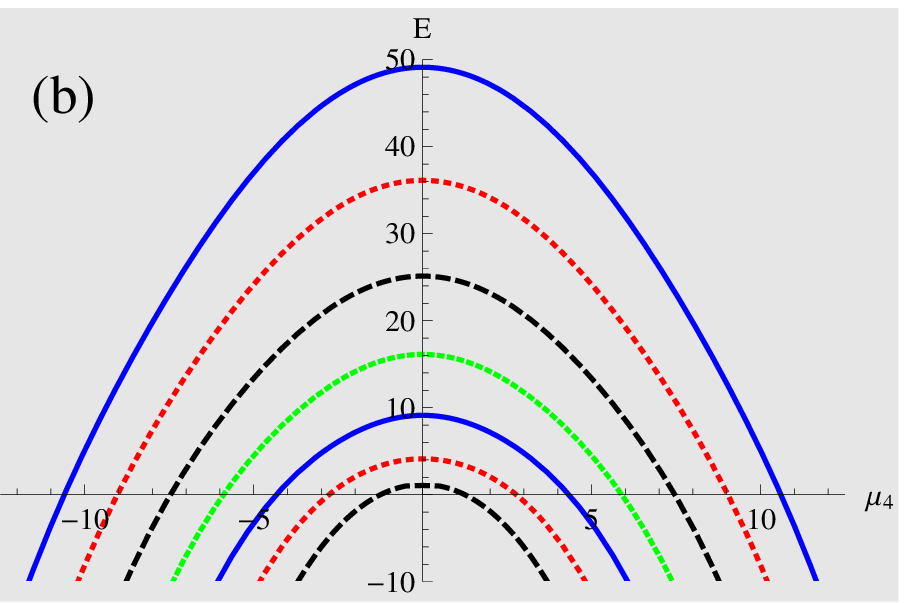} %
\includegraphics[width=7.5cm,height=6.0cm]{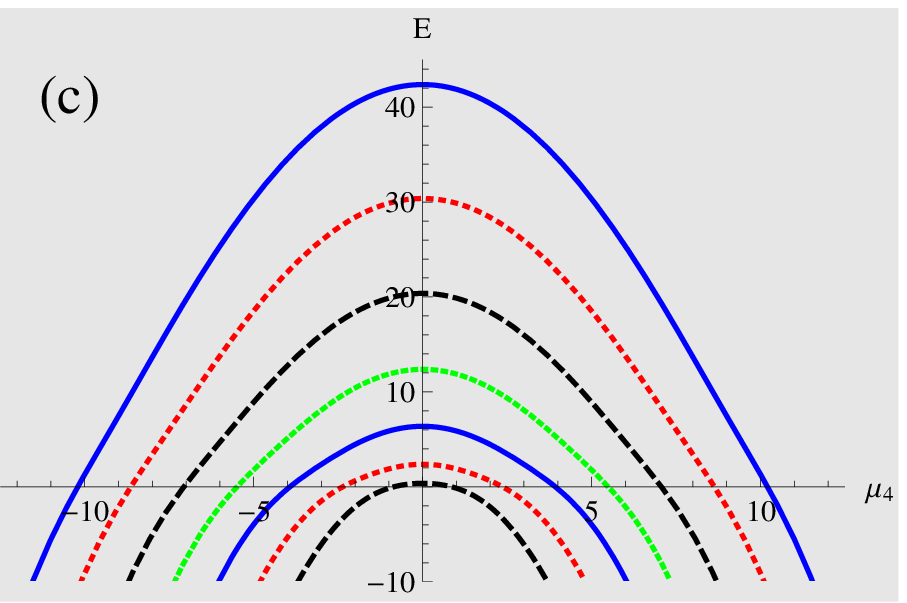} %
\includegraphics[width=7.5cm,height=6.0cm]{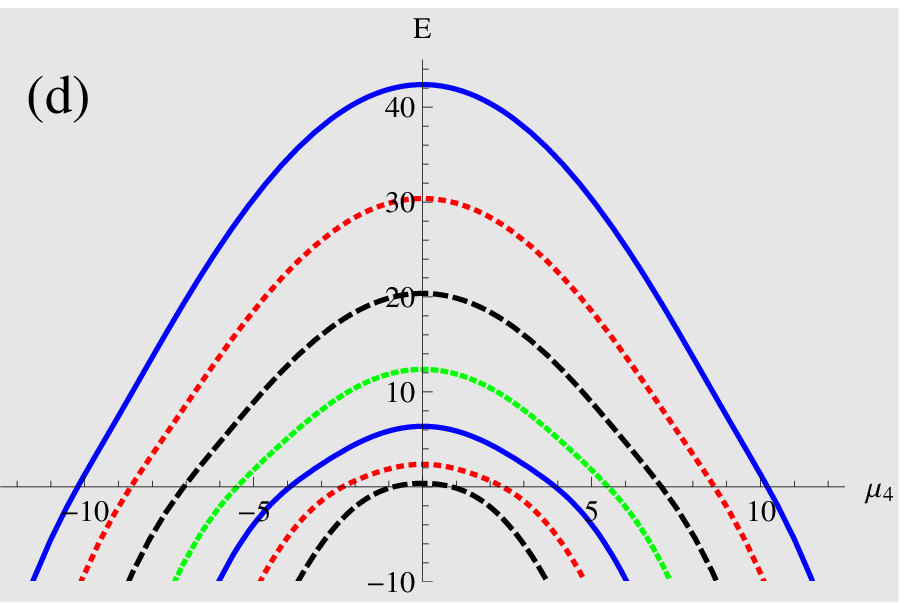} \centering   
\caption{Entirely real energy eigenvalue spectrum for the non-Hermitian
Hamiltonian $H_{\mathcal{PT}_{5}}^{(3)}$ as a function of $\protect\mu _{4}$
with $\protect\mu _{3}=1/2$ and $\protect\mu _{7}=0$. The values for even
(odd) eigenfunctions with bosonic and fermionic boundary conditions are
displayed in the panels a and c (b and d), respectively.}
\label{fig1}
\end{figure}

We commence our numerical analysis by taking $\mu _{7}=0$. In this case the
map $\eta $ is well-defined, except when $\mu _{3}=\mu _{4}$ for which $%
\lambda \rightarrow \infty $ by (\ref{cons2}). Thus we expect an entirely
real energy spectrum. In figure \ref{fig1} we present the results of our
numerical solutions for the computation of the lowest seven energy levels,
demonstrating that this is indeed the case for the even and odd solutions
for bosonic as well as fermionic boundary conditions.

\begin{figure}[h]
\centering   \includegraphics[width=7.5cm,height=6.0cm]{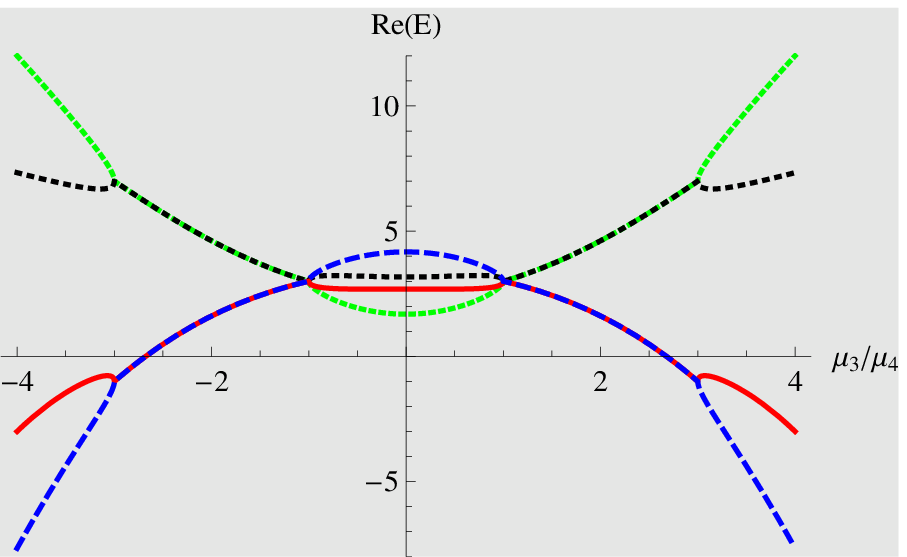} %
\includegraphics[width=7.5cm,height=6.0cm]{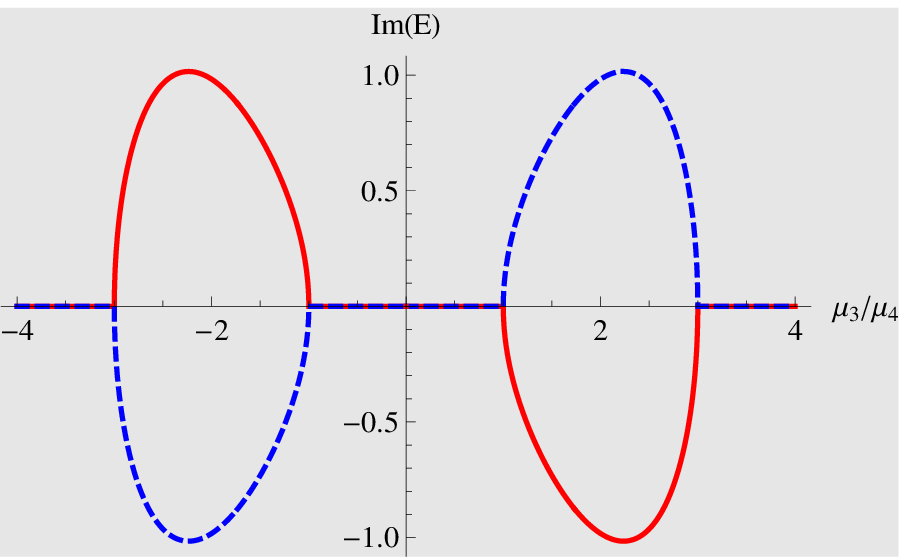} \centering   
\caption{Spontaneously broken energy eigenvalue spectra for $H_{\mathcal{PT}%
_{5}}^{(3)}$ as a function of $\protect\mu_3$ with fixed values $\protect\mu%
_4=1$ and $\protect\mu_7=4$ with even (green, short dashed) and odd (black,
dotted) eigenfunctions for bosonic boundary conditions and as a function of $%
\protect\mu_4$ with fixed values $\protect\mu_3=1$ and $\protect\mu_7=4$
with even (red, solid) and odd (blue, dashed) eigenfunctions for bosonic
boundary conditions. The exceptional points are located at ($\protect\mu%
_{3/4}=\pm 1, E=3$), ($\protect\mu_{3}=\pm 3, E=7$) and ($\protect\mu%
_{4}=\pm 3, E=-1$).}
\label{fig2}
\end{figure}

For nonzero values of $\mu _{7}$ we can enter the ill-defined region for the
Dyson map as for the last constraint in (\ref{cons2}) we may encounter
values on the right hand side between $-1$ and $1$. Viewing the energy
eigenvalues as functions of $\mu _{3/4}$ we expect therefore to find four
exceptional points at $\mu _{3/4}=\pm \mu _{4/3}\pm \sqrt{\mu _{7}}$. As an
example we fix $\mu _{3/4}=1$ and $\mu _{7}=4$, such that $\eta (\mu _{4/3})$
is only well defined for $|\mu _{4/3}|<1$ or $|\mu _{4/3}|>3$. Indeed our
numerical solutions for this choice presented in figure \ref{fig2} confirm
this prediction. We observe that the eigenvalues acquire a complex part when 
$1<\mu _{3/4}<3$ and $-3<\mu _{3/4}<-1$ and is real otherwise. We present
here only the spectrum for bosonic boundary condition with an even
wavefunction since the qualitative behaviour for the other cases and levels
are very similar as already noted in the previous example.

We clearly observe the typical behaviour of spontaneously broken $\mathcal{PT%
}$-symmetry in form of two of the real eigenvalues merging into complex
conjugate pairs at exceptional points. We further note that there are three
disconnected regions $|\mu _{3/4}|<1$ or $|\mu _{3/4}|>3$ in which all the
eigenvalues are real.

\begin{figure}[h]
\centering   \includegraphics[width=7.5cm,height=6.0cm]{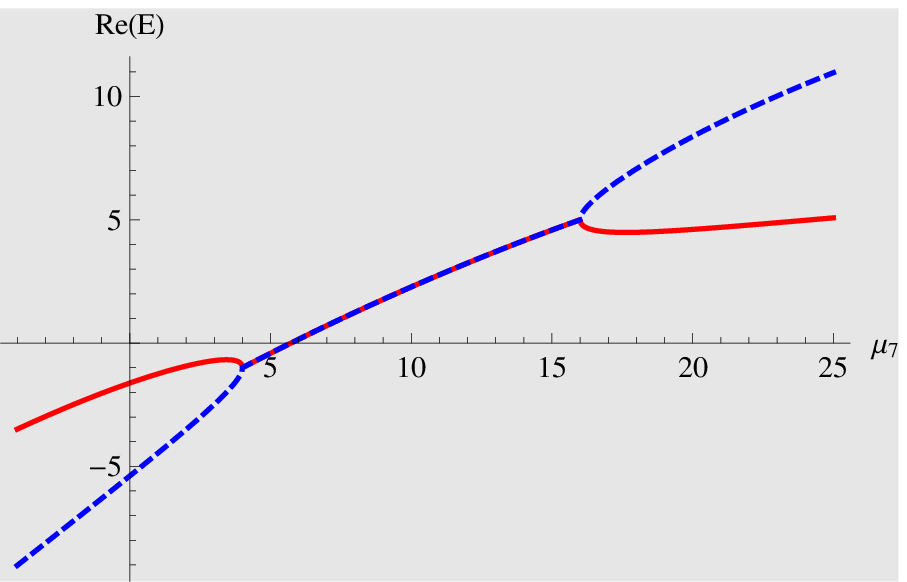} %
\includegraphics[width=7.5cm,height=6.0cm]{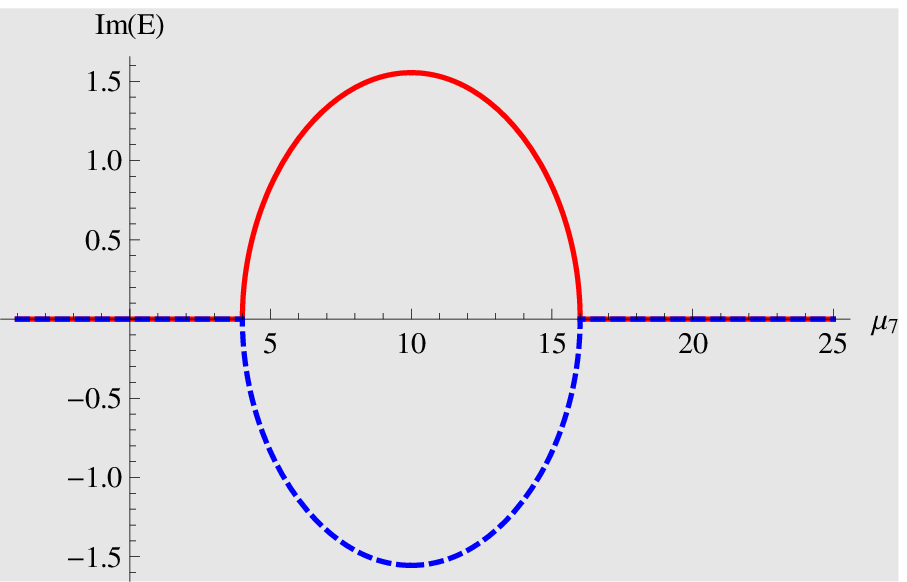} \centering   
\caption{Spontaneously broken energy eigenvalue spectra for $H_{\mathcal{PT}%
_{5}}^{(3)}$ as a function of $\protect\mu_7$ with fixed values $\protect\mu%
_3=1$ and $\protect\mu_4=3$ with even (red, solid) and odd (blue, dashed)
eigenfunctions. The exceptional points are located at ($\protect\mu_{7}=4 ,
E=-1$) and ($\protect\mu_{7}=16, E=5$).}
\label{fig3}
\end{figure}

Alternatively we may also view the energy spectra as functions of $\mu _{7}$%
, in which case we expect just two exceptional points at $(\mu _{3}\pm \mu
_{4})$ $^{2}$. Our numerical solutions for this choice are presented in
figure \ref{fig3}, which clearly confirms these values and the predicted
qualitative behaviour.

\begin{figure}[h]
\centering   %
\includegraphics[width=7.5cm,height=6.0cm]{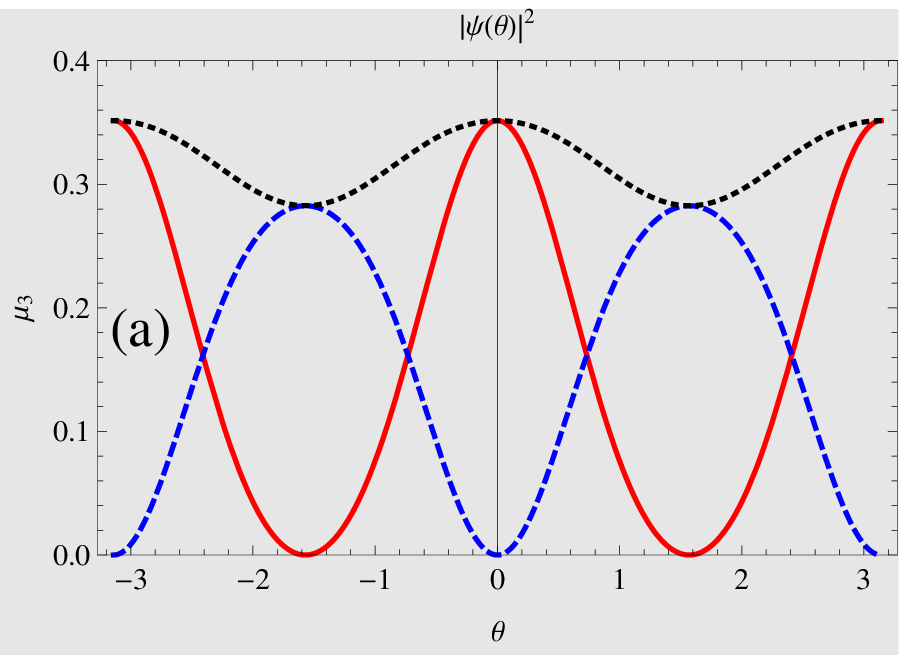} %
\includegraphics[width=7.5cm,height=6.0cm]{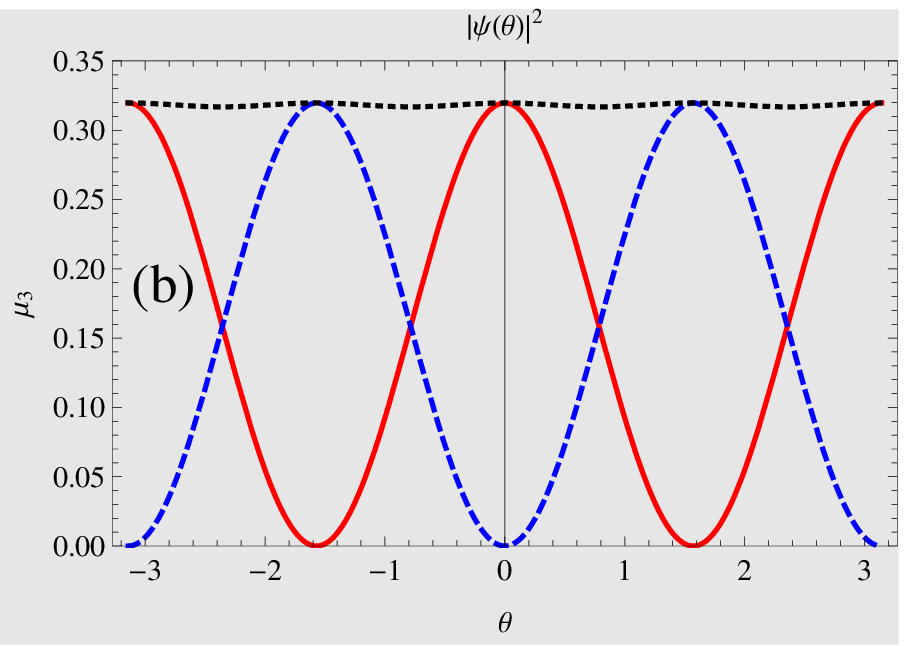} 
\centering   
\caption{Intensities for a merging an even (red, solid) and odd (blue,
dashed) wavefunction together with their sum (black, dotted) in the unbroken
with $\protect\mu _{3}=0.8$, $\protect\mu _{4}=1$, $\protect\mu _{7}=4$ and
broken $\mathcal{PT}$-regime with $\protect\mu _{3}=1.2$, $\protect\mu %
_{4}=1 $, $\protect\mu _{7}=4$, panel (a) and (b), respectively.}
\label{figInt}
\end{figure}

We conclude this subsection by considering the intensities, as in principle
these quantities are experimentally accessible. In figure \ref{figInt} we
display the intensity $I(\theta )=|\psi (\theta )|^{2}$ for an odd and even
wavefunction merging at the exceptional points whose energy spectrum is
displayed in figure \ref{fig2}. In the spontaneously broken $\mathcal{PT}$%
-regime we clearly observe the loss/gain symmetry around the line $I_{\max
}(\theta )/2$, which is absent in the unbroken $\mathcal{PT}$-regime.

\begin{figure}[h]
\centering   \includegraphics[width=12.5cm]{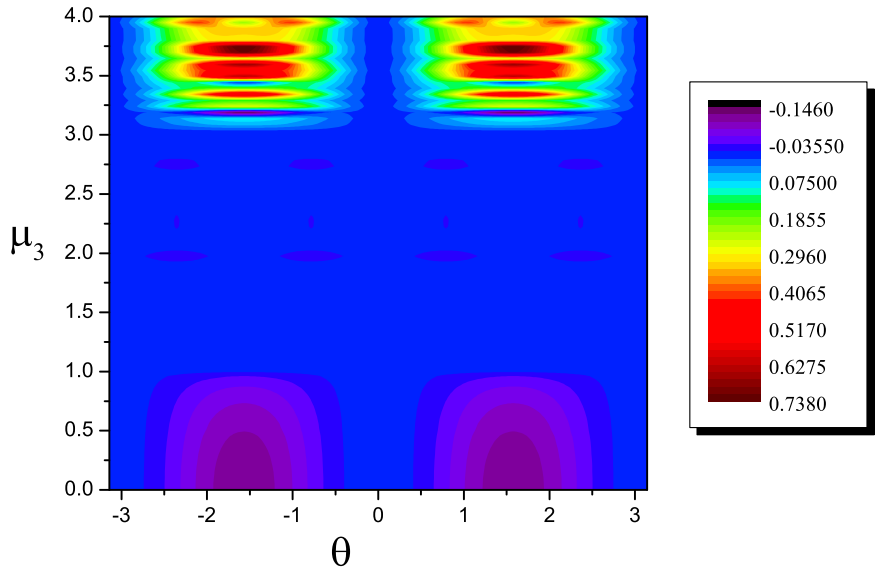} \centering   
\caption{Intensity sum $I(\protect\theta )=|\protect\psi _{\text{even}}(%
\protect\theta )|^{2}+|\protect\psi _{\text{odd}}(\protect\theta )|^{2}-|%
\protect\psi _{\text{even}}(0)|^{2}$ as a function of $\protect\mu _{3}$
with fixed values $\protect\mu _{4}=1$ and $\protect\mu _{7}=4$. }
\label{figp}
\end{figure}

In figure \ref{figp} we scan over a larger range for the coupling constant $%
\mu _{3}$ entering and leaving the broken $\mathcal{PT}$-regime and depict
the sum $I(\theta )=|\psi _{\text{even}}(\theta )|^{2}+|\psi _{\text{odd}%
}(\theta )|^{2}-|\psi _{\text{even}}(0)|^{2}$. We clearly observe an
oscillatory behaviour in the unbroken $\mathcal{PT}$-regime ($\mu _{3}<1$
and $\mu _{3}>3$) and complete annihilation in the region where the symmetry
is spontaneously broken ($1$ $<\mu _{3}<3$). This qualitative behaviour is
reminiscent of the symmetric gain/loss behaviour observed in complex optical
potentials \cite{Guo}.

\subsubsection{Sinusoidal optical lattices}

For different choices we can also make contact with a simpler example
currently of great interest, since it can be realized experimentally in form
of optical lattices. Making the simple choice%
\begin{equation}
\mu _{1}=1,\quad \mu _{2}=\mu _{3}=\mu _{4}=\mu _{5}=\mu _{6}=0\quad \tau
=\rho ,\quad \coth (2\lambda )=\frac{\mu _{7}-\mu _{8}}{\mu _{9}},
\label{conop}
\end{equation}%
we obtain the isospectral Hermitian counterpart%
\begin{equation}
h_{\mathcal{PT}_{4/5}}^{(ol)}=J^{2}+\frac{1}{2}\sqrt{(\mu _{7}-\mu
_{8})^{2}-\mu _{9}^{2}}(v^{2}-u^{2})+\frac{1}{2}(\mu _{7}+\mu _{8}).
\label{OP}
\end{equation}%
Taking the representation (\ref{rep1}) in (\ref{OP}), the further special
choices $\mu _{7}=0$, $\mu _{8}=-4$, $\mu _{9}=-8V_{0}$ or $\mu _{7}=-\mu
_{8}=A/2$, $\mu _{9}=-2AV_{0}$ reduce the potential to the sinusoidal
optical lattice potential dealt with in \cite{OPMidya} or \cite{MatHugh},
respectively. In both cases the requirement for the validity of the Dyson
map $\left\vert (\mu _{7}-\mu _{8})/\mu _{9}\right\vert <1$, implied by the
last equation in (\ref{conop}), boils down to $\left\vert V_{0}\right\vert
<1/2$ confirming the finding in \cite{OPMidya} and \cite{MatHugh} that only
in this regime the corresponding potential leads to a real energy eigenvalue
spectrum.

\subsubsection{Complex Mathieu equation}

\begin{figure}[h]
\centering   \includegraphics[width=7.5cm,height=6.0cm]{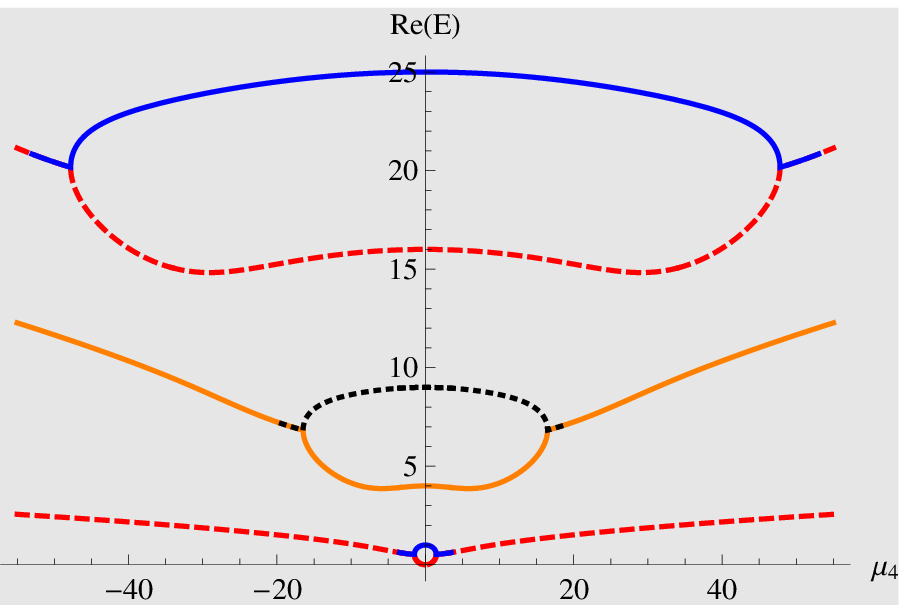} %
\includegraphics[width=7.5cm,height=6.0cm]{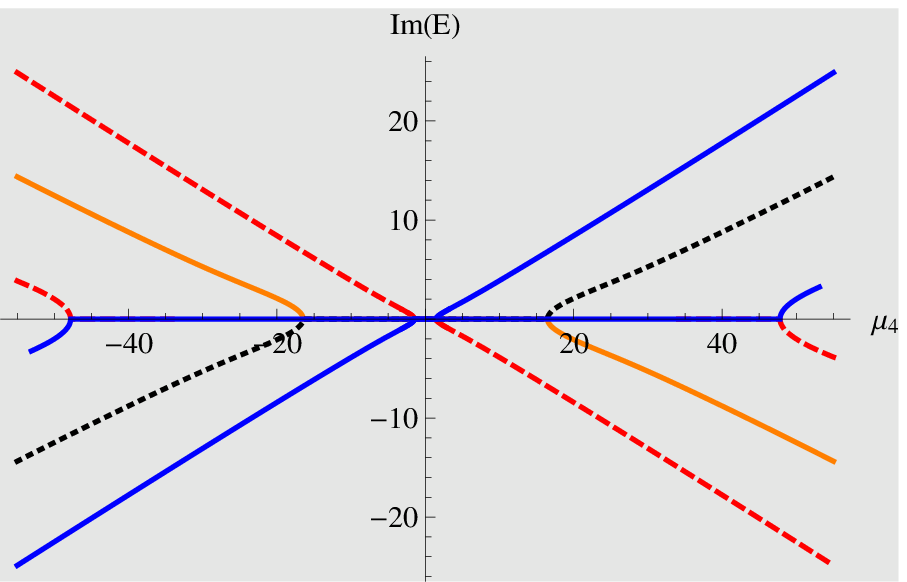} \centering   
\caption{Spontaneously broken energy eigenvalue spectra for the parameter
choice (2.37) as a function of $\protect\mu_4$ with even eigenfunctions for
bosonic boundary conditions. The exceptional points are located at $(\protect%
\mu_4=\pm 1.4687, E=0.5205)$, $(\protect\mu_4=\pm 16.47116, E=6.8323)$ and $(%
\protect\mu_4=\pm 47.80596, E=20.1677)$.}
\label{fig4}
\end{figure}

We conclude by discussing the parameter choice%
\begin{equation}
\mu _{1}=1,\quad \mu _{2}=0,\quad \mu _{3}=-\frac{\mu _{6}}{2},\quad \mu
_{5}=-\mu _{4},\quad \mu _{7}=\frac{\mu _{4}^{2}}{2},\quad \mu _{8}=-\frac{%
\mu _{6}^{2}}{4},\quad \mu _{9}=-\frac{\mu _{4}\mu _{6}}{2}.
\end{equation}%
In that case the reported similarity transformation is invalid. However,
similarly as in the previous case we may solve the corresponding Schr\"{o}%
dinger equation exactly by mapping it to the Mathieu equation, which is
however complex in this case. We then find the solution 
\begin{equation}
\psi (\theta )=e^{-i\mu _{4}/2\cos \theta +\mu _{6}/2\sin \theta }\left[
c_{1}C\left( 4E,i\mu _{4},\theta /2\right) +c_{2}S\left( 4E,i\mu _{4},\theta
/2\right) \right] .  \label{sol3}
\end{equation}%
As in the previous case we impose bosonic or fermionic boundary conditions
to determine the spectrum. Our results are depicted in figure \ref{fig4}.

We clearly observe the usual merger of two energy levels at the exceptional
points where they split into complex conjugate pairs. Since the real part of
the energy eigenvalues is monotonically increasing we note that the spectrum
is entirely real for $\left\vert \mu _{4}\right\vert \leq 1.46876$. It
remains an open challenge to explain the origin of this value for instance
by finding an exact similarity transformation. As we expect, this behaviour
is similar to the one reported in \cite{BenKal}.

\section{$\mathcal{PT}$-symmetric E$_{3}$-invariant systems}

The $E_{3}$-algebra is the rank 3 extension of the $E_{2}$-algebra, spanned
by six generators $J_{i}$, $P_{i}$ for $i=1,2,3$ satisfying the algebra%
\begin{equation}
\left[ J_{j},J_{k}\right] =i\varepsilon _{jkl}J_{l},\qquad \left[ J_{j},P_{k}%
\right] =i\varepsilon _{jkl}P_{l},\qquad \text{and\qquad }\left[ P_{j},P_{k}%
\right] =0.
\end{equation}%
Evidently every subset $\{J_{j},P_{k},P_{l}\}$ with $j\neq k\neq l$
constitutes an $E_{2}$-subalgebra. It is convenient to introduce the
following combinations of the generators%
\begin{equation}
J_{z}=2J_{1},\quad J_{\pm }=J_{2}\pm iJ_{3},\quad P_{z}=P_{1},\qquad \text{%
and\qquad }P_{\pm }=\pm P_{2}+iP_{3},
\end{equation}%
such that we obtain the commutation relations 
\begin{equation}
\left[ J_{z},J_{\pm }\right] =\pm 2J_{\pm },~~\left[ J_{+},J_{-}\right]
=J_{z},~~\left[ J_{z},P_{\pm }\right] =\pm 2P_{\pm },~~\left[ J_{\pm },P_{z}%
\right] =-P_{\pm },~~\left[ J_{\pm },P_{\mp }\right] =-2P_{z},  \label{E3pm}
\end{equation}%
with all remaining ones vanishing. In \cite{Douge3} the following
representation was provided for this algebra%
\begin{equation}
\begin{array}{lll}
J_{z}:=x\partial _{x}-y\partial _{y},~~\quad & J_{+}:=x\partial _{y},~~ & 
J_{-}:=y\partial _{x}, \\ 
P_{z}:=-xy\partial _{z},~~ & P_{+}:=x^{2}\partial _{z},~\quad & 
P_{-}:=y^{2}\partial _{z}.%
\end{array}
\label{rep}
\end{equation}%
Similarly as $E_{2}$, also $E_{3}$ is left invariant with respect to various
types of $\mathcal{PT}$-symmetries%
\begin{equation}
\begin{array}{llllll}
\mathcal{PT}_{1}:~~~~~ & J_{k}\rightarrow -J_{k}, & P_{k}\rightarrow -P_{k},
& i\rightarrow -i; &  &  \\ 
\mathcal{PT}_{2}: & J_{k}\rightarrow -J_{k}, & P_{k}\rightarrow P_{k}, & 
i\rightarrow -i; &  &  \\ 
\mathcal{PT}_{3}: & J_{k}\rightarrow J_{k}, & P_{1}\rightarrow P_{1}, & 
P_{2}\leftrightarrow P_{3}, & i\rightarrow -i; &  \\ 
\mathcal{PT}_{4}: & J_{1}\rightarrow -J_{1},~~~~ & J_{2/3}\rightarrow
J_{2/3},~~~~ & P_{1/3}\leftrightarrow -P_{1/3},~~~ & P_{2}\leftrightarrow
P_{2},~ & i\rightarrow -i;%
\end{array}%
\end{equation}%
for $k=1,2,3$.

Once again we wish to find the Dyson map to map non-Hermitian Hamiltonians
expressed in terms of bilinear combinations of these generators to Hermitian
ones. For the $E_{3}$-algebra we take it to be of the general form%
\begin{equation}
\eta =e^{\lambda _{z}J_{z}+\lambda _{+}J_{+}+\lambda _{-}J_{-}+\kappa
_{z}P_{z}+\kappa _{+}P_{+}+\kappa _{-}P_{-}},\qquad \text{\ \ \ \ \ \ \ \
for }\lambda _{z},\lambda _{\pm },\kappa _{z},\kappa _{\pm }\in \mathbb{R}.
\label{eta2}
\end{equation}%
For the adjoint action of this operator on the $E_{3}$-generators we compute%
\begin{equation}
\eta P_{\ell }\eta ^{-1}=\mu _{\ell z}P_{z}+\mu _{\ell +}P_{+}+\mu _{\ell
-}P_{-}\qquad \text{for }\ell =z,\pm
\end{equation}%
with constant coefficients 
\begin{eqnarray*}
\mu _{zz} &=&1+2c(\omega )\lambda _{+}\lambda _{-},\quad \mu _{\pm \pm
}=1+(2\lambda _{z}^{2}+\lambda _{+}\lambda _{-})c(\omega )\pm 2s(\omega
)\lambda _{z},\quad \\
\mu _{\pm \mp } &=&c(\omega )\lambda _{\mp }^{2},\quad \mu _{\pm z}=\mp
2c(\omega )\lambda _{z}\lambda _{\mp }-2s(\omega )\lambda _{\mp },\quad \mu
_{z\pm }=\mp c(\omega )\lambda _{z}\lambda _{\pm }-s(\omega )\lambda _{\pm },
\end{eqnarray*}%
and 
\begin{equation}
\eta J_{\ell }\eta ^{-1}=\nu _{\ell z}J_{z}+\nu _{\ell +}J_{+}+\nu _{\ell
-}J_{-}+\rho _{\ell z}P_{z}+\rho _{\ell +}P_{+}+\rho _{\ell -}P_{-}\qquad 
\text{for }\ell =z,\pm
\end{equation}%
with constant coefficients%
\begin{eqnarray*}
\nu _{zz} &=&1+2c(\omega )\lambda _{+}\lambda _{-},\quad \nu _{\pm \pm }=1+%
\tilde{\omega}^{2}c(\omega )\pm 2s(\omega )\lambda _{z},\quad \nu _{\pm \mp
}=-c(\omega )\lambda _{\mp }^{2}, \\
\nu _{\pm z} &=&\mp s(\omega )\lambda _{\mp }-c(\omega )\lambda _{z}\lambda
_{\mp },\quad \nu _{z\pm }=-2c(\omega )\lambda _{z}\lambda _{\pm }\mp
2s(\omega )\lambda _{\pm },
\end{eqnarray*}%
\begin{eqnarray*}
\rho _{zz} &=&4\left[ \left( \lambda _{-}\kappa _{+}-\lambda _{+}\kappa
_{-}\right) c(\omega )-\frac{\lambda _{+}\lambda _{-}}{\omega ^{2}}\mu
(c(\omega )-s(\omega ))\right] \\
\rho _{z\pm } &=&c(\omega )(\pm \lambda _{\pm }\kappa _{z}-2\lambda
_{z}\kappa _{\pm })\mp 2s(\omega )(\kappa _{\pm }+\lambda _{\pm }\kappa
_{z})\pm \frac{2c(\omega )}{\omega ^{2}}\lambda _{\pm }\nu +\frac{s(\omega )%
}{\omega ^{2}}\lambda _{\pm }\left( \mu \mp 2\nu \right) \\
&&-\frac{\cosh (2\omega )}{\omega ^{2}}\mu \lambda _{\pm } \\
\rho _{\pm z} &=&c(\omega )(\lambda _{\mp }\kappa _{z}\pm 2\lambda
_{z}\kappa _{\mp })+2s(\omega )(\kappa _{\mp }-\lambda _{\mp }\kappa _{z})+%
\frac{2c(\omega )}{\omega ^{2}}\lambda _{\mp }\nu \pm \frac{s(\omega )}{%
\omega ^{2}}\lambda _{\mp }\left( \mu \mp 2\nu \right) \\
&&\mp \frac{\cosh (2\omega )}{\omega ^{2}}\mu \lambda _{\mp } \\
\rho _{\pm \pm } &=&\pm c(\omega )\tilde{\mu}+s(\omega )\kappa _{z}\pm \mu 
\frac{\tilde{\omega}^{2}}{\omega ^{2}}[s(\omega )-c(\omega )]+\frac{\cosh
(2\omega )-s(\omega )}{\omega ^{2}}\lambda _{z}\mu \\
\rho _{\pm \mp } &=&-2c(\omega )\lambda _{\mp }\kappa _{\mp }\pm \frac{\mu
\lambda _{\mp }^{2}}{\omega ^{2}}[s(\omega )-c(\omega )]
\end{eqnarray*}%
where we abbreviated $\omega :=\sqrt{\lambda _{z}^{2}+\lambda _{+}\lambda
_{-}}$, $\tilde{\omega}:=\sqrt{2\lambda _{z}^{2}+\lambda _{+}\lambda _{-}}$, 
$\mu :=\kappa _{z}\lambda _{z}+\kappa _{+}\lambda _{-}-\kappa _{-}\lambda
_{+}$, $\tilde{\mu}:=2\kappa _{z}\lambda _{z}+\kappa _{+}\lambda _{-}-\kappa
_{-}\lambda _{+}$, $\nu :=\kappa _{+}\lambda _{z}\lambda _{-}-\kappa
_{z}\lambda _{+}\lambda _{-}-\kappa _{-}\lambda _{z}\lambda _{+}$, $c(\omega
):=(\cosh (2\omega )-1)/(2\omega ^{2})$ and $s(\omega ):=\sinh (2\omega
)/(2\omega )$.

The construction of isospectral counterparts, if they exist, for
non-Hermitian Hamiltonians symmetric with regard to various different types
of $\mathcal{PT}$-symmetries is far more involved in this for this algebra.
The most generic cases are very complicated in this case as they involve 25
free parameters. One may therefore restrict the discussion to simpler
examples, such as for instance the complements of $E_{2}$ in $E_{3}$
constitutes well-defined subclasses

For instance, we may consider a $\mathcal{PT}_{1}$-invariant Hamiltonians of 
$E_{3}/E_{2}$-Lie algebraic type. Selecting $\{J_{z},P_{\pm }\}$ as the
generators of the $E_{2}$-subalgebra the most general Hamiltonian of this
type is 
\begin{equation}
\tilde{H}_{\mathcal{PT}_{1}}=\mu _{1}J_{+}^{2}+\mu _{2}J_{-}^{2}+\mu
_{3}P_{z}^{2}+\mu _{4}P_{z}J_{+}+\mu _{5}P_{z}J_{-}+\mu _{6}J_{+}J_{-}+i\mu
_{7}J_{+}+i\mu _{8}J_{-}+i\mu _{9}P_{z}.
\end{equation}%
All the necessary tools have been provided here to find the corresponding
counterparts etc. We leave this discussion for future investigations \cite%
{DFMprep}.

\section{Conclusion}

We presented five different types of $\mathcal{PT}$-symmetries (\ref{PT})
for the Euclidean algebra $E_{2}$ (\ref{E2}). Considering the most general
invariant non-Hermitian Hamiltonians in terms of bilinear combinations of
the generators of this algebra, we have systematically constructed
isospectral counterparts from Dyson maps $\eta $ of the general form (\ref%
{eta}) by exploiting its adjoint action on the Lie algebraic generators. In
this process some of the coupling constants involved had to be constrained.
We noted that the different versions of the symmetries also lead to
qualitatively quite different isospectral counterparts. For the symmetries $%
\mathcal{PT}_{1}$ and $\mathcal{PT}_{2}$ the required constraints rendered
the original Hamiltonians $H_{\mathcal{PT}_{1/2}}$ Hermitian, such that the
adjoint action of $\eta $ maps Hermitian Hamiltonians to Hermitian ones. It
should be noted that the maps are non-trivial, albeit the distinguishing
features of the obtained Hamiltonians $h_{\mathcal{PT}_{1/2}}$ remain
unclear. More interesting are the transformation properties of the
non-Hermitian Hamiltonians invariant under the symmetries $\mathcal{PT}_{3}$%
, $\mathcal{PT}_{4}$ and $\mathcal{PT}_{5}$, as they lead to genuine
non-Hermitian/Hermitian isospectral pairs constructed from an explicit
non-perturbative Dyson map.

For the representation (\ref{rep1}) we analyzed the $\mathcal{PT}_{5}$%
-system in further detail by solving the corresponding time-dependent Schr%
\"{o}dinger equation. For some parameter choices we found simple
transformations of the real Mathieu equation as solutions. In a subset of
cases the corresponding energy spectra were identified to be entirely real,
see figure \ref{fig1}. For other choices we observed spontaneously broken $%
\mathcal{PT}$-symmetry with region in the parameter space where the whole
spectrum remained real. It is possible to consider the spectra as functions
of coupling constants in such a way that its monotonic variation leads to an
initial break down of the $\mathcal{PT}$-symmetry at some exceptional points
which is subsequently regained, see figure \ref{fig2}. This numerically
observed behaviour is completely understood from the explicit formulae for
the Dyson maps, which break down at the exceptional points. In section
2.5.2. we have made contact to some simple systems of optical lattices and
it should be highly interesting to investigate further whether the more
involved systems with richer structure we considered here may also be
realized experimentally. We have verified the typical gain/loss symmetry for
one of those models.

Clearly we have not exhausted the discussion for the entire parameter space
for the $\mathcal{PT}_{5}$-system and also left the analysis of
time-dependent Schr\"{o}dinger equation $\mathcal{PT}_{3}$ and $\mathcal{PT}%
_{4}$ for further investigation. An additional open problem is the analysis
of alternative representations such as (\ref{rep2}) and many more not
mentioned here. Also still an intriguing open challenge is the computation
of the explicit Dyson map for systems of the type dealt with in section
2.5.3. We established that they certainly require a different type of Ansatz
for the Dyson map $\eta $ as the one in (\ref{eta}).

The completion of the above mentioned programme is far from being finished
for the Euclidean algebra $E_{3}$. For that case we have provided the far
more complicated adjoint action on the generators and left the further
analysis, which can be carried out along the same lines as for $E_{2}$, for
future investigations \cite{DFMprep}. \medskip

\noindent \textbf{Acknowledgments:} SD is supported by a City University
Research Fellowship. TM is funded by an Erasmus Mundus scholarship and
thanks City University for kind hospitality.

\newif\ifabfull\abfulltrue


\end{document}